\documentclass[twocolumn,english,superscriptaddress,prl,amsmath,amssymb,floatfix,longbibliography]{revtex4-1}
\usepackage[T1]{fontenc}
\usepackage[utf8]{inputenc}
\usepackage{color}
\usepackage{babel}
\usepackage{amsmath}
\usepackage{graphicx}
\usepackage{esint}
\usepackage[unicode=true,pdfusetitle,
 bookmarks=true,bookmarksnumbered=false,bookmarksopen=false,
 breaklinks=false,pdfborder={0 0 0},pdfborderstyle={},backref=false,colorlinks=true]
 {hyperref}

\makeatletter

\usepackage{dcolumn}
\usepackage{bm}
\usepackage{color}
\usepackage{soul}
\usepackage{babel}
\usepackage{amsfonts}
\usepackage{slashed}
\usepackage{enumerate}

\usepackage{babel}

\makeatother

\begin{document}
\global\long\def\sgn{\mathrm{sgn}}%
\global\long\def\ket#1{\left|#1\right\rangle }%
\global\long\def\bra#1{\left\langle #1\right|}%
\global\long\def\sp#1#2{\langle#1|#2\rangle}%
\global\long\def\abs#1{\left|#1\right|}%
\global\long\def\avg#1{\langle#1\rangle}%

\title{Engineering two-qubit mixed states with weak measurements}
\author{Parveen Kumar}
\affiliation{Department of Condensed Matter Physics, Weizmann Institute of Science,
Rehovot 76100, Israel}
\author{Kyrylo Snizhko}
\affiliation{Department of Condensed Matter Physics, Weizmann Institute of Science,
Rehovot 76100, Israel}
\author{Yuval Gefen}
\affiliation{Department of Condensed Matter Physics, Weizmann Institute of Science,
Rehovot 76100, Israel}
\begin{abstract}
It is known that protocols based on weak measurements can be used
to steer quantum systems into predesignated pure states. Here we show
that weak-measurement-based steering protocols can be harnessed for
on-demand engineering of \emph{mixed} states. In particular, through
a continuous variation of the protocol parameters, one can guide a
classical target state to a discorded one, and further on, towards
an entangled target state.
\end{abstract}
\maketitle
\emph{Introduction}.---A generalized quantum measurement comprises
a two-step protocol: (i) switching on, and later off, an interaction
Hamiltonian, coupling the quantum system and the quantum detector,
leading to a unitary evolution of the combined setup over a prescribed
time interval; (ii) performing a projective measurement of the decoupled
detector, which leads to a probabilistic quantum jump \citep{VonNeumannJohn2018,Wiseman2010a,Hubert2014}.
The detector readout provides information about the system’s state.
Measurements are designated strong (projective) or weak, based on
the system-detector interaction strength. The former collapses the
system to one of the eigenstates of the measured observable. By contrast,
generalized (a.k.a. weak) measurements may result in a slight nudge
to the system state \citep{Arthurs1965,Aharonov1988a,Brun2002a,Svensson2013,Patel2017}.
No matter how weak the measurement is, it always creates an unavoidable
impact on the system state through its back-action \citep{Korotkov1999,Korotkov2001}.
Traditionally, this measurement-induced back-action was considered
an undesirable effect since the primary purpose of a measurement is
to extract information about the system without perturbing it.

Following a disparate paradigm, one may employ the measurement-induced
back-action on the system's state as a means to control the system's
evolution, steering \citep{SteeringFoot} it towards a predesignated
pure target state \citep{Pechen2006,Roa2006,Roa2007,Jacobs2010,Ashhab2010}.
A recent work (cf.~\citep{Roy2019} and references therein) analyzed
a host of protocols utilizing \emph{“blind”} measurements \citep{BlindMeasFoot}
for engineering pure target states in single- and many-body systems.

Traditionally, mixed states are not the objective of quantum steering.
In practice, non-pure states may appear due to errors or noise of
steering protocols. In a paradigmatic shift, we put them here as the
target of steering, emphasizing the interest in and the value of mixed
states. Their quantumness can be expressed through their discord \cite{Henderson2001,Ollivier2002,Modi2012,Hunt2019,*Kumar2020c,*Hunt2020}.
Discorded quantum states have been proposed as resources for various
quantum information tasks: achieving quantum speedup \citep{Datta2008,Lanyon2008},
remote state preparation \citep{Dakic2012}, and quantum purification
protocols \citep{Bennett1996,Murao1998}. Engineering predesignated
mixed states is, therefore, a task of interest.

The present analysis introduces a measurement-based protocol which
can be used to steer a two-qubit system to an arbitrary predesignated
state (pure or mixed), independently of the system's initial state
(the latter is assumed unknown). We illustrate our protocol by considering
a family of target states. These, depending on the protocol parameters,
may be: (i) non-discorded and non-entangled (``classical''), (ii)
discorded and non-entangled, or (iii) discorded and entangled, thus
providing us with a smooth navigation tool from classical to fully
quantum states.

The guiding principle of our protocol is as follows. Under sufficiently
weak measurement, the system evolution can be described by a Lindbladian
master equation. Such an equation will have at least one (and in the
present case exactly one) steady state that is approached exponentially
quickly. We design the protocol such that this steady state is the
target state. We first recall a measurement-based protocol whose steady
state is an arbitrarily chosen pure state \citep{Roy2019}. In order
to generate a mixed state, we diagonalize the density matrix of the
target state and juxtapose the protocols for stabilizing each of the
density matrix's pure eigenstates. In the explicit example we consider,
we find that the rate of converging towards the target state does
not significantly depend on its discord or degree of entanglement,
so that our protocol is equally efficient independently of the target
state's usefulness for quantum information purposes.
\begin{figure}
\begin{centering}
\includegraphics[width=1\columnwidth]{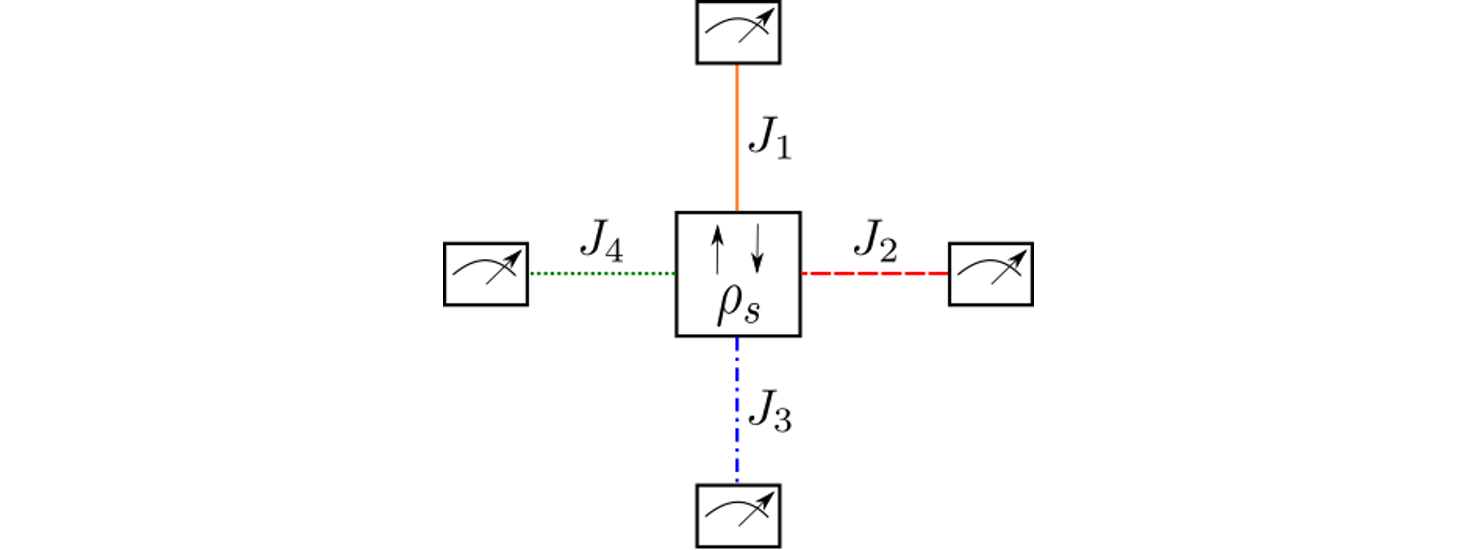}
\par\end{centering}
\caption{Setup for steering a two-qubit system to an arbitrary mixed state.
Four detectors are coupled to the system, couplings $J_{1}$, $J_{2}$,
$J_{3}$, and $J_{4}$ are switched on and off in turns such that
only one of them is non-zero at any given time. Repeated measurement
by any single detector would steer the system towards a pure state
$|B_{i}\rangle\langle B_{i}|$, $i=1,...,4$. With all the detectors
operating alternatingly, the system is steered towards a mixed state
of the form (\ref{eq:two_qubit_rho_target}). \label{fig:schematic_diagram}}
\end{figure}

\emph{General evolution under repeated blind measurements.}---Consider
a quantum system in state represented by the density matrix $\rho_{s}$
and a quantum detector prepared in state $\rho_{d}^{(0)}$. Before
they interact, the joint system-detector state can be written as

\begin{equation}
\rho(t)=\rho_{s}(t)\otimes\rho_{d}^{(0)}.\label{eq:total_state_of_composite_system}
\end{equation}
In order to perform a measurement, the system interacts with the detector
via an interaction Hamiltonian $H_{s-d}$; the joint system-detector
state evolves as

\begin{equation}
\rho(t+\tau)=U\rho(t)U^{\dagger},\label{eq:composite_state_evoln}
\end{equation}
where $U=\exp\left(-iH_{s-d}\tau\right)$, and $\tau$ is the interaction
time. Subsequently the detector state is measured projectively, disentangling
the composite system-detector state and generating a measurement back-action
on the system state. When discarding (i.e., tracing out) the measurement
readouts, a procedure we denote ``blind measurement'', the effect
of the back-action is represented through

\begin{equation}
\rho_{s}(t+\tau)=\text{Tr}_{d}\left[\rho(t+\tau)\right].\label{eq:system_state_only_gen}
\end{equation}
Following each measurement step, the detector is reset to its initial
state $\rho_{d}^{(0)}$, and then the same measurement procedure is
repeated. This protocol gives rise to a non-trivial evolution of the
system state.

Denoting the system-detector interaction time $\tau$, and taking
the continuous time limit $\tau\rightarrow dt$, one arrives at the
following differential equation for the system state evolution under
blind measurements

\begin{multline}
\frac{d\rho_{s}}{dt}=\mathcal{L}[\rho_{s}]=i\text{Tr}_{d}\left(\left[\rho(t),H_{s-d}\right]\right)\\
-\frac{1}{2}\text{Tr}_{d}\left(\left[H_{s-d},\left[H_{s-d},\rho(t)\right]\right]\tau\right).\label{eq:composite_system_evoln_eq}
\end{multline}
Here $\mathcal{L}$ is the Liouvillian superoperator acting on the
system state, and we dropped the terms $O(\tau^{2})$ on the r.h.s.
The first term on the right-hand-side of the above equation generates
unitary evolution of $\rho_{s}$, while the second term represents
dissipative evolution and can be cast in the form of a Lindbladian.
In other words, the above equation can be written as

\begin{multline}
\frac{d\rho_{s}}{dt}=\mathcal{L}[\rho_{s}]\\
=i[\rho_{s}(t),H_{s}]-\frac{1}{2}\sum_{j}\left(\{L_{j}^{\dagger}L_{j},\rho_{s}(t)\}-2L_{j}\rho_{s}(t)L_{j}^{\dagger}\right),\label{eq:gen_master_eq_for_system}
\end{multline}
where $\{\cdot,\cdot\}$ represents the anti-commutator, $H_{s}$
is the effective system Hamiltonian, and $L_{j}$ are the Lindblad
jump operators acting on the system state. This way, the sequence
of measurements influences the system state in a quasicontinuous manner
and ultimately steers it to a steady state determined by the condition

\begin{equation}
\frac{d\rho_{s}^{(T)}}{dt}=\mathcal{L}[\rho_{s}^{(T)}]=0.\label{eq:steady_state_soln_eq}
\end{equation}

\emph{Steering towards a pure target state.---}We now recall the
principles of measurement-based steering protocol of Ref.~\citep{Roy2019},
focusing on the two-qubit case, at the center of our analysis. The
protocol facilitates stabilizing the system in an arbitrary pure target
state $\ket{B_{1}}$, corresponding to $\rho_{s}^{(T)}=|B_{1}\rangle\langle B_{1}|$.

To implement the protocol we first select three arbitrary states $\ket{B_{2}}$,
$\ket{B_{3}}$, and $\ket{B_{4}}$, such that together with $\ket{B_{1}}$
they form an orthonormal basis in the four-dimensional Hilbert space
of the two-qubit system. Three steps now follow: in the $k^{\text{th}}$
step, $k=1,2,3$, a measurement is performed with a system-detector
coupling that is designed such that the measurement back-action steers
the system away from $\ket{B_{k+1}}$. This is accomplished by choosing
$H_{s-d}^{k}=J\left(|B_{1}\rangle\langle B_{k+1}|\otimes\sigma^{-}+h.c.\right)$,
where the detector is a single qubit acted upon by the Pauli matrices
$\sigma^{\pm}=\left(\sigma_{x}^{(d)}\pm i\sigma_{y}^{(d)}\right)/2$,
and $J$ is the coupling strength (for simplicity it is the same in
all the three steps). Before each measurement, the detector is initialized
in $\rho_{d}^{(0)}=\ket{\uparrow}\bra{\uparrow}$. With the duration
of each step being $\tau$, the density matrix evolution over $dt=3\tau$
is given by

\begin{equation}
\frac{d\rho_{s}}{dt}=-\frac{g}{2}\sum_{j=1}^{4}\left(\{L_{j}^{\dagger}L_{j},\rho_{s}\}-2L_{j}\rho_{s}L_{j}^{\dagger}\right),\label{eq:master_eq_with_H1_pure_state_stab}
\end{equation}
where $g=J^{2}\tau/3$, and the jump operators $L_{j}=(1-\delta_{1j})|B_{1}\rangle\langle B_{j}|$
(we introduced $L_{1}=0$ for completeness). It follows from Eq.~(\ref{eq:master_eq_with_H1_pure_state_stab})
that

\begin{equation}
\frac{d\rho_{s}}{dt}=0\Leftrightarrow\rho_{s}=\rho_{s}^{(T)}=|B_{1}\rangle\langle B_{1}|,\label{eq:b1b1_steady_state}
\end{equation}
i.e., the system is steered towards the desired pure state.

It is instructive to understand the mechanism by which the above steering
works. In each step, the system-detector interaction Hamiltonian $H_{s-d}^{k}$
steers the system in a two-dimensional subspace spanned by $|B_{1}\rangle$
and $|B_{k+1}\rangle$ from the state $|B_{k+1}\rangle$ to $|B_{1}\rangle$
without affecting the rest of $|B_{i}\rangle$ $(i\neq k+1)$. Since
$H_{s-d}^{k}$ commutes with $|B_{1}\rangle\langle B_{1}|\otimes\rho_{d}^{(0)}$,
the measurement does not disturb the system if it is in state $\ket{B_{1}}$,
and the detector then remains in its initial state, $|\uparrow\rangle$.
This makes $\ket{B_{1}}$ not just a steady state of the evolution,
but a dark state in the terminology of Ref.~\citep{Diehl2008}: once
the system is in $\ket{B_{1}}$ it is not affected by the detectors.
If the system is in $\ket{B_{k+1}}$, a transition to $\ket{B_{1}}$
(accompanied by the detector state flipping to $\ket{\downarrow}$)
happens with probability $\sin^{2}J\tau$. Likewise, with probability
$\cos^{2}J\tau$, the detector does not flip the state and the system
remains in $\ket{B_{k+1}}$. Note that if the system is initially
in a (coherent or incoherent) superposition of $\ket{B_{1}}$ and
$\ket{B_{k+1}}$, both detector readouts affect the system state:
$\ket{\downarrow}$ state of the detector implies the system has jumped
from $\ket{B_{k+1}}$ to $\ket{B_{1}}$, while $\ket{\uparrow}$ implies
a change of the weights of the superposition due to different probabilities
of the $\uparrow$ readout depending on the system state (this is
referred in the literature as a ``null weak measurement'' \citep{Ruskov2007,Zilberberg2013a,Zilberberg2014a,Gebhart2020}
or by a number of different names \citep{Elitzur2001,Paraoanu2006,Xu2011}).
Averaging over the possible detector readouts and taking the limit
$J\tau\ll1$, one obtains the master equation (\ref{eq:master_eq_with_H1_pure_state_stab}).

\begin{figure}
\begin{centering}
\includegraphics[width=1\columnwidth]{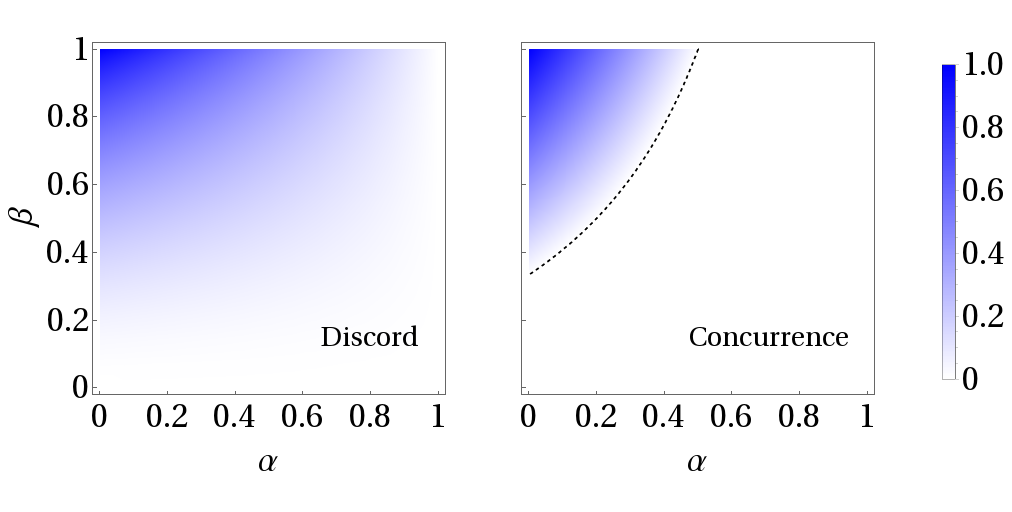}
\par\end{centering}
\caption{Quantum discord (\ref{eq:quantum_discord_example_state}) and concurrence
(\ref{eq:concurrence_for_example_state}) for the two-qubit state
$\tilde{\rho}$ of Eq.~(\ref{eq:two_qubit_example_state}) as a function
of the parameters $\alpha$ and $\beta$. The black dashed line, $\alpha=\frac{3\beta-1}{3\beta+1}$,
separates the regions of zero and non-zero concurrence. The discord
is non-zero everywhere except for the lines $\alpha=1$ and $\beta=0$.\label{fig:discord_and_concurrence_for_the_exanple_state}}
\end{figure}

\emph{Steering towards a mixed target state.---}We now focus on the
key result of the paper: steering the two-qubit system to a desired
mixed state. Any mixed target state $\rho_{s}^{(T)}$ has a spectral
decomposition \citep{Nielsen2010}

\begin{equation}
\rho_{s}^{(T)}=\sum_{i=1}^{4}p_{i}|B_{i}\rangle\langle B_{i}|,\label{eq:two_qubit_rho_target}
\end{equation}
where $\ket{B_{i=1,...,4}}$ form an orthonormal basis in the two-qubit
Hilbert space, and $p_{i}\geq0$ is the probability of the system
being in the corresponding $\ket{B_{i}}$ state, so that $\sum_{i=1}^{4}p_{i}=1$.
The protocol described in the previous section can be used to steer
the system to $\ket{B_{1}}$. Furthermore, by exchanging the roles
of $\ket{B_{1}}$ and one of $\ket{B_{i\neq1}}$, the protocol steers
the system to the corresponding $\ket{B_{i}}$. We now show that combining
the four protocols, each steering the system to one of $\ket{B_{i}}$,
with appropriate coupling strengths, $J\rightarrow J_{i}$, allows
to stabilize the mixed state in Eq.~(\ref{eq:two_qubit_rho_target}).

A schematic experimental setup for this complex protocol is presented
in Fig.~\ref{fig:schematic_diagram}. Each part of the protocol,
steering the system towards one of the $\ket{B_{i}}$ lasts $3\tau$.
Consequently, the density matrix evolution for $dt=4\times3\tau=12\tau$
is described by

\begin{eqnarray}
\frac{d\rho_{s}}{dt} & = & \mathcal{L}[\rho_{s}]=\sum_{i=1}^{4}\mathcal{L}_{i}[\rho_{s}]\nonumber \\
 & = & -\frac{1}{2}\sum_{i=1}^{4}g_{i}\sum_{j=1}^{4}\biggl(\{L_{j}^{(i)\dagger}L_{j}^{(i)},\rho_{s}\}-2L_{j}^{(i)}\rho_{s}L_{j}^{(i)\dagger}\biggr),\label{eq:master_equation_two_qubits}
\end{eqnarray}
where $g_{i}=J_{i}^{2}\tau/12$ and $L_{j}^{(i)}=(1-\delta_{ij})|B_{i}\rangle\langle B_{j}|$.
Equation (\ref{eq:master_equation_two_qubits}) has a unique steady
state,

\begin{equation}
\rho_{s}=\rho_{s}^{(T)}=\frac{1}{\sum_{i=1}^{4}g_{i}}\biggl(\sum_{j=1}^{4}g_{j}|B_{j}\rangle\langle B_{j}|\biggr).\label{eq:two_qubit_steady_state_final}
\end{equation}
 We note that choosing

\begin{equation}
g_{j}=\bar{g}\,p_{j}\,,\label{eq:g_for_two_qubit_rho_target}
\end{equation}
where $j=1,2,3,4$ and $\bar{g}=\sum_{i=1}^{4}g_{i}$, stabilizes
the desired target state in Eq.~(\ref{eq:two_qubit_rho_target}).
Therefore, measurement-based steering towards an arbitrary mixed state
requires diagonalizing the density matrix of the latter, bringing
the state to the form (\ref{eq:two_qubit_rho_target}), and concurrent
utilization of the four protocols, each stabilizing one of the pure
eigenstates which make up the mixed-state density matrix.

We emphasize that the simple-looking result of Eq.~(\ref{eq:g_for_two_qubit_rho_target})
is highly non-trivial. Indeed, for arbitrary four Lindbladians $\tilde{\mathcal{L}}_{i}$
such that $\tilde{\mathcal{L}}_{i}[|B_{i}\rangle\langle B_{i}|]=0$,
one cannot guarantee that $\sum_{i}g_{i}\tilde{\mathcal{L}}_{i}[\sum_{j}g_{j}|B_{j}\rangle\langle B_{j}|]=\sum_{i,j}g_{i}g_{j}\tilde{\mathcal{L}}_{i}[|B_{j}\rangle\langle B_{j}|]=0$.
The fact that this holds in our case is a special feature of the measurements
we employ here (i.e., of the Lindbladians describing the measurement-induced
evolution). Employing other types of measurements may also allow for
stabilizing arbitrary states, however, the relation between $g_{j}$
and $p_{j}$ may not be as simple as Eq.~(\ref{eq:g_for_two_qubit_rho_target}).

If we record the detector's readouts, the dynamics of the system subject
to our protocol can be understood as follows: The limit of weak measurement
addressed here, $J_{i}\tau\ll1$, implies that a detector click (one
of the detectors flipping its state to $\ket{\downarrow}$ during
the measurement) is a rare event. It follows that the system, initially
prepared in a possibly coherent superposition of different $\ket{B_{i}}$,
will experience coherence-preserving dynamics induced by a ``no-click''
back-action for a long time $\sim\bar{g}^{-1}$ (this is akin to ``null
weak measurements'' \citep{Ruskov2007,Zilberberg2013a,Zilberberg2014a,Gebhart2020}).
Eventually, a click will register, bringing the system to one of the
$\ket{B_{i}}$ states and destroying coherence between different $\ket{B_{i}}$
components. From this point on, the no-click dynamics does not affect
the system state, while rare clicks make it jump between different
$\ket{B_{i}}$ states. The system spending random amounts of time
in different $\ket{B_{i}}$ results in the average state given by
Eq.~(\ref{eq:two_qubit_rho_target}). Notably, the system, while
being \emph{on average} in the target state, probabilistically jumps
among the constituent pure states $\ket{B_{i}}$, which is manifested
by occasional detector clicks. In other words, a mixed target state
in our protocol is a steady state but not a dark state.

\begin{figure}
\begin{centering}
\includegraphics[width=1\columnwidth]{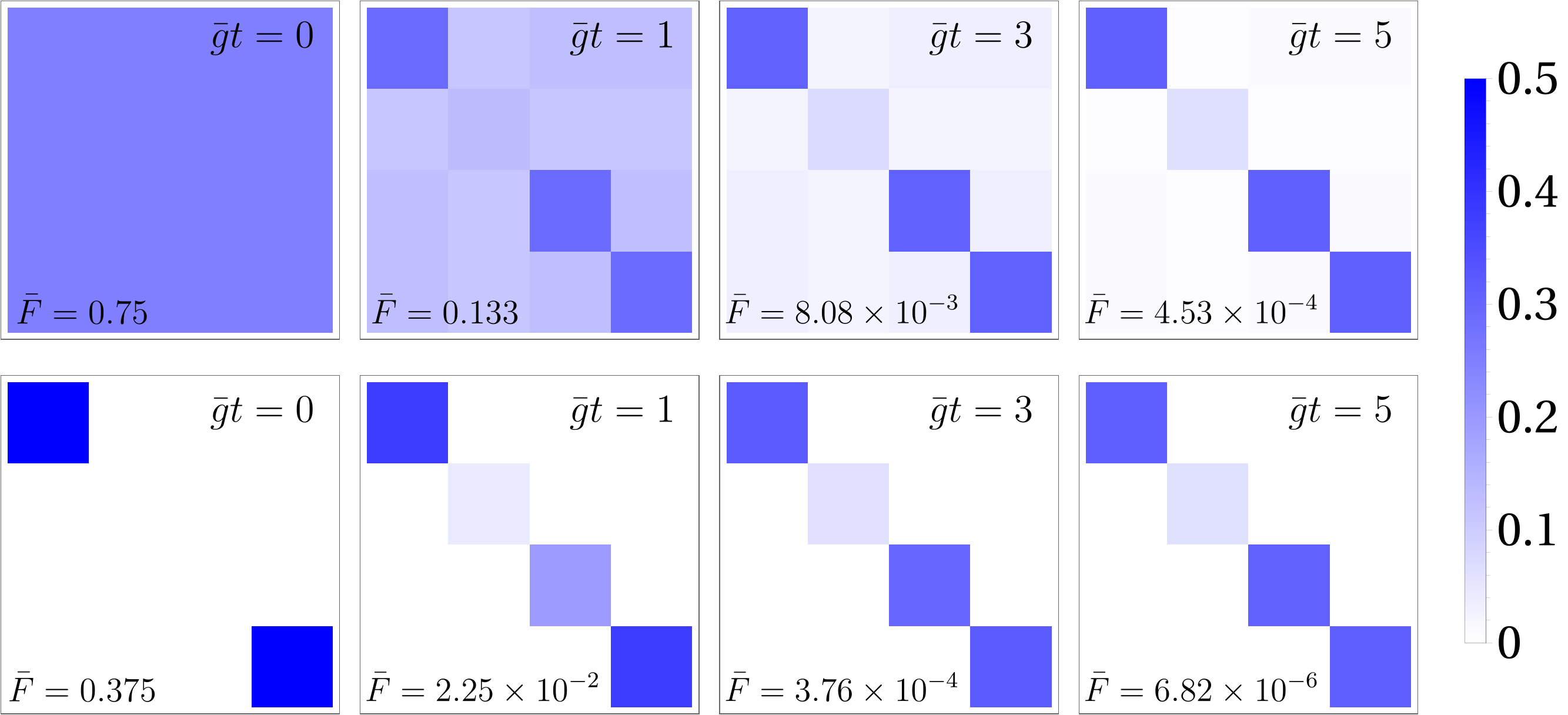}
\par\end{centering}
\caption{The density matrix of the two-qubit system as it is steered towards
a target state, $\tilde{\rho}$, of Eq.~(\ref{eq:two_qubit_example_state}).
The magnitudes of the density matrix entries in the basis $\protect\ket{B_{i}}$
(\ref{eq:basis_example_state}) are represented by a color scale.
The initial state can be pure ($\rho_{s}(0)=\frac{1}{4}\sum_{i,j}|B_{i}\rangle\langle B_{j}|$,
top) or mixed ($\rho_{s}(0)=\frac{1}{2}\left(|B_{1}\rangle\langle B_{1}|+|B_{4}\rangle\langle B_{4}|\right)$,
bottom). $\bar{F}=1-\left(\text{Tr\ensuremath{\sqrt{\sqrt{\rho(t)}\rho^{(T)}\sqrt{\rho(t)}}}}\right)^{2}$
is the deviation from perfect fidelity of the target state preparation.
Irrespective of the initial state, the system has essentially converged
to the target state with $\alpha=\beta=\frac{1}{2}$ at $\bar{g}t=5$.
\label{fig:time_evoln_of_the_example_state_steering}}
\end{figure}

\emph{Steering to ``classical'' vs ``quantum'' states}.---We
now illustrate our protocol with an example where the system can be
steered into a family of states which can be classical or quantum
depending on the measurement couplings employed. Consider the following
two-qubit state,

\begin{equation}
\tilde{\rho}=p_{1}|\uparrow\uparrow\rangle\langle\uparrow\uparrow|+p_{2}|\psi^{+}\rangle\langle\psi^{+}|+p_{3}|\psi^{-}\rangle\langle\psi^{-}|+p_{4}|\downarrow\downarrow\rangle\langle\downarrow\downarrow|,\label{eq:two_qubit_example_state}
\end{equation}
where $|\psi^{\pm}\rangle=\frac{1}{\sqrt{2}}\left(|\uparrow\downarrow\rangle\pm|\downarrow\uparrow\rangle\right),$
and

\begin{eqnarray}
p_{1}=\frac{(1-\beta+\alpha(1+\beta))}{4} & , & \quad p_{2}=\frac{(1-\alpha)(1-\beta)}{4},\nonumber \\
p_{3}=\frac{(1-\alpha)(1+3\beta)}{4} & , & \quad p_{4}=\frac{(1-\beta+\alpha(1+\beta))}{4}.\label{eq:example_state_eigenvals}
\end{eqnarray}
Here $\alpha$ and $\beta$ are two independent parameters such that
$0\le\alpha,\beta\le1$, and the coefficients $p_{i}$ correspond
to the $p_{i}$ in Eq.~(\ref{eq:two_qubit_rho_target}). This state
may or may not have quantum correlations depending on $\alpha$ and
$\beta$. For example, using Peres-Horodecki criterion \citep{Peres1996,Horodecki1996},
one shows that $\tilde{\rho}$ is separable (not entangled) if and
only if $\alpha\geq\frac{3\beta-1}{3\beta+1}$. 

Quantum correlations are commonly quantified via concurrence (a measure
of entanglement) \citep{Wootters1998,Modi2012} and quantum discord
\citep{Henderson2001,Ollivier2002,Modi2012,Hunt2019}. Calculating
discord $\mathcal{Q}$ and concurrence $\mathcal{C}$ for an arbitrary
state is a challenging task, but for the state in Eq.~(\ref{eq:two_qubit_example_state})
both of them can be calculated analytically \citep{Wootters1998,Luo2008,Ali2010,Modi2012}.
They are respectively given by

\begin{eqnarray}
\mathcal{Q}(\tilde{\rho}) & = & \frac{1-\alpha}{4}\biggl((1-\beta)\log_{2}[(1-\alpha)(1-\beta)]\nonumber \\
 & - & 2(1+\beta)\log_{2}[(1-\alpha)(1+\beta)]\nonumber \\
 & + & (1+3\beta)\log_{2}[(1-\alpha)(1+3\beta)]\biggr),\label{eq:quantum_discord_example_state}
\end{eqnarray}
and

\begin{equation}
\mathcal{C}(\tilde{\rho})=\begin{cases}
\frac{3\beta(1-\alpha)-(1+\alpha)}{2}\textrm{\ensuremath{\quad\text{for}}} & \alpha<\frac{3\beta-1}{3\beta+1},\\
0 & \text{otherwise}.
\end{cases}\label{eq:concurrence_for_example_state}
\end{equation}
Note that the discord $\mathcal{Q}(\tilde{\rho})$ is only zero when:
(i) $\alpha=1$ or (ii) $\beta=0$, while the concurrence $\mathcal{C}(\tilde{\rho})$
exhibits a sharp change of behavior at finite $\alpha$ and $\beta$,
cf.~Fig.~\ref{fig:discord_and_concurrence_for_the_exanple_state}.
The state $\tilde{\rho}$ can thus be purely \emph{classical} (both
discord and concurrence vanish), \emph{discorded} (concurrence vanishes),
or \emph{entangled} (concurrence and discord are both non-zero) depending
on the parameters $\alpha$ and $\beta$.

\begin{figure}
\begin{centering}
\includegraphics[width=1\columnwidth]{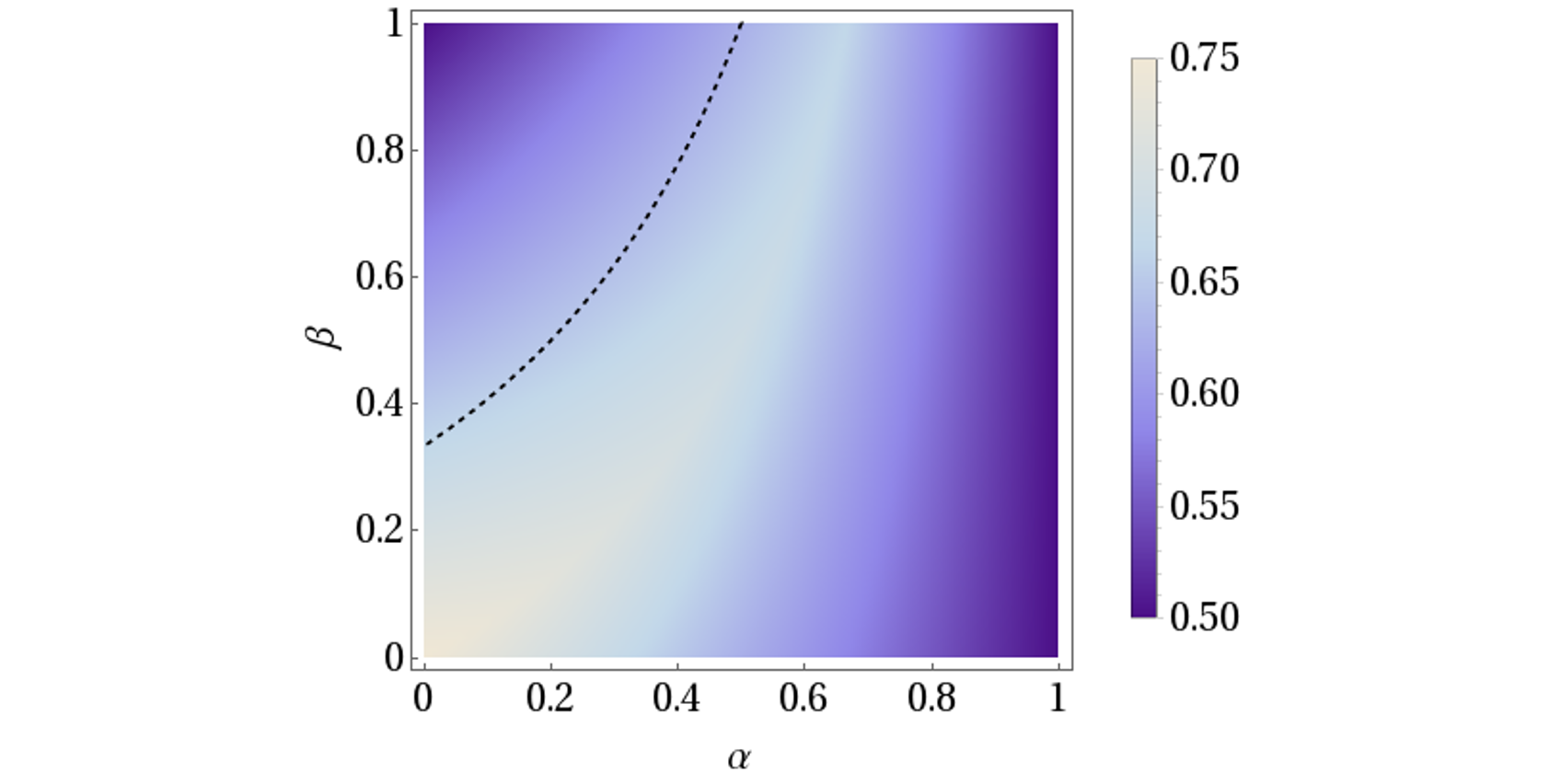}
\par\end{centering}
\caption{The convergence rate in units of $\bar{g}$ as a function of parameters
$\alpha$ and $\beta$. The black dashed line separates the regions
of zero and non-zero concurrence, cf.~Fig.~\ref{fig:discord_and_concurrence_for_the_exanple_state}.
The convergence rate does not depend significantly on the target state.
In particular, it does not depend on whether the target state is classical,
discorded, or entangled.\label{fig:steering-rate-for-different-target-state}}
\end{figure}

We may generate $\tilde{\rho}$ using the protocol described above.
First, the target state density matrix should be diagonalized. It
is evident from Eq.~(\ref{eq:two_qubit_example_state}) that the
eigenbasis of $\tilde{\rho}$ is

\begin{equation}
|B_{1}\rangle=|\uparrow\uparrow\rangle,~|B_{2}\rangle=|\psi^{+}\rangle,~|B_{3}\rangle=|\psi^{-}\rangle,~|B_{4}\rangle=|\downarrow\downarrow\rangle.\label{eq:basis_example_state}
\end{equation}
Using Eq.~(\ref{eq:g_for_two_qubit_rho_target}), one obtains that
the couplings $g_{i}$ in Eq.~(\ref{eq:master_equation_two_qubits})
are $g_{i}=\bar{g}\,p_{i},$ where $\bar{g}=\sum_{i}g_{i}$ characterizes
the total strength of all measurements employed. Figure~\ref{fig:time_evoln_of_the_example_state_steering}
illustrates the time evolution of the two-qubit system as it is steered
from an initial state to the target state $\tilde{\rho}$. Deviations
from the target state decay exponentially in time; the decay rates
are determined by the real parts of the non-zero eigenvalues of the
Liouvillian superoperator $\mathcal{L}$, cf.~Eq.~(\ref{eq:master_equation_two_qubits}).
The smallest of the decay rates is referred to as the convergence
rate; it is determined by the inverse of the Liouvillian gap---the
smallest in magnitude real part of all non-zero eigenvalues. The dependence
of the convergence rate on the target state is presented in Fig.~\ref{fig:steering-rate-for-different-target-state}.
Note that the dependence on $\alpha$ and $\beta$ is not too strong,
implying that our protocol works equally well for both entangled and
non-entangled states, as well as for states in the vicinity of the
transition between the two regions.

\emph{Discussion}.---We have proposed a measurement-based protocol
that can generate any two-qubit state by design (pure or mixed), starting
from an arbitrary unknown initial state. We illustrate the protocol
with an example, in which the target state can be classical, discorded,
or entangled, depending on relative strengths of the measurements
employed in the protocol.

Note that conventional studies of open system evolution (e.g., \citep{Gao1997,Prosen2011,Albert2014})
are concerned with finding the steady state of a given Liouvillian.
By contrast, here we are concerned with stabilizing a predesignated
state (\textquotedbl target state\textquotedbl ), i.e., finding
the evolution protocol (Liouvillian) for which our target state is
the steady state. We also note that our approach is distinct from
drive-and-dissipation schemes \citep{Wu2007,Diehl2008,Kraus2008,Roncaglia2010,Pechen2011,Diehl2011,Murch2012,Bardyn2013,Leghtas2013,Liu2016,Goldman2016,Lu2017,Huang2018},
where environment is employed to relax the system to a desired state.
The two crucial distinctions here are: (a) the relaxation is induced
by measurements, implying the possibility to use the measurement readouts
to confirm the system's desired behavior (and, possibly, hasten the
convergence towards the target state); (b) our system does not have
a Hamiltonian (no ``drive'').

Our protocol can be implemented in a variety of experimental platforms.
The main ingredient of our protocol, blind measurement stabilizing
the system at a specific pure state, is particularly natural for implementation
in cold-ion systems \citep{Av2020}. The nature of cold-ion experiments
does not allow for registering individual measurement readouts in
order to verify the protocol behavior. This can, however, be done
by working with superconducting ``artificial atoms'' \citep{Vijay2011,Minev2019}.

An interesting future direction concerns stabilizing $N$-qubit mixed
states. Stabilizing $N$-qubit pure states in an efficient manner
is already a challenging problem, which, however, has been solved
for the restricted class of tensor-network states \citep{Kraus2008,Verstraete2009,Roy2019}.
Stabilizing $N$-qubit mixed states turns out to be an even more challenging
problem still under research \citep{Bondarenko2018}. Our two-qubit
protocol may be a useful block in future $N$-qubit protocols.

\begin{acknowledgments}

We acknowledge funding by the Deutsche Forschungsgemeinschaft (DFG,
German Research Foundation) -- Projektnummer 277101999 -- TRR 183 (project
C01) and Projektnummer EG 96/13-1, by the Italia-Israel project QUANTRA,
by the Israel Science Foundation (ISF), and by the National Science
Foundation--Binational Science Foundation (NSF-BSF) EAGER program.

\end{acknowledgments}

\bibliography{references,extra}

\begin{thebibliography}{63}
\bibitem [{Ste()}]{SteeringFoot}%
  \BibitemOpen
  \href@noop {} {}\bibinfo {note} {This is not to be confused with the term
  "steering" from the field of quantum infromation, which defines a special
  class of correlations. In this paper, we use the word in a completely
  different sense, "steering" for us is the name of the process that takes an
  arbitrary initial state and leads it to a predesignated target
  state.}\BibitemShut {Stop}%
\bibitem [{Bli()}]{BlindMeasFoot}%
  \BibitemOpen
  \href@noop {} {}\bibinfo {note} {This term refers to a measurement where the
  final state of the detector (the "readout") is traced out, discarding the
  acquirement of information through the measurement process but accounting for
  measurement back-action.}\BibitemShut {Stop}%
\end{thebibliography}

\end{document}